\documentclass[11pt]{article}
\usepackage{cospar}
\usepackage{url}
\def\lapprox{\;\rlap{\lower 2.5pt            
    \hbox{$\sim$}}\raise 1.5pt\hbox{$<$}\;} 

\usepackage{graphics}
\usepackage[figuresright]{rotating}

\title{MICRO-EVENTS IN THE ACTIVE AND QUIET SOLAR CORONA}
\author{A. O. Benz and P. C. Grigis \vskip-1.5cm \address{Institute of Astronomy, ETH, 8092 Zurich, Switzerland}}

\begin{document}

\maketitle

\begin{abstract}
The content of hot material in the corona is not constant. Soft X-ray and high-temperature EUV line observations show that new material, apparently heated and evaporated from the chromosphere, is frequently injected into the corona both in active and quiet regions. {\sl Active regions} are found to exhibit transient brightenings, termed here microflares, due to such enhancements in emission measure. They appear at a rate of up to 10 per hour in RHESSI observations of 3--15 keV X-rays, occurring even during the periods of lowest solar activity so far in the mission. The RHESSI observations combined with measurements at other wavelengths yield estimates of the energy input into the corona. These observations suggest that the models for coronal heating must be complemented with respect to continuous replenishing the lower corona by chromospheric material heated to coronal temperatures. The observed micro-events are secondary phenomena and do not represent the primary energy release, nor its total amount. Nevertheless, they are an interesting source of information on the heating process(es) of the corona. The micro-events are compared to events in {\sl quiet regions}, termed here nanoflares, which seem to be a different population, well separated in temperature and emission measure from microflares. 
\end{abstract}

\section*{INTRODUCTION}
\vskip3mm
The heating of the solar corona has been a riddle since the discovery of the high coronal temperature in the late 1930's. There are elaborate investigations and many proposals on possible sources of energy (e.g. review by Mandrini {\sl et al.}, 2000; and recent original work by Galsgaard \& Nordlund 1996; Hood, Ireland \& Priest 1997; Erdelyi 1998; Sturrock 1999; Sakai, Takahata, \& Sokolov 2001), but proof can only come from observations of the corona. What are observable signatures of the heating process? A spatial relation between coronal emission (thus heating) and magnetic flux has been reported (Schrijver, 1987; Fisher {\sl et al.}, 1998; Fludra \& Ireland, 2003). Of particular interest are time variations in coronal temperature, density and energy content, as they may reveal clues on the nature of any non-stationary heating process. New instruments have made it possible to study this question with sufficient resolution.

Heating the corona by flare-like events was first proposed by Gold (1957). The terms "microflare" and "nanoflare" have been introduced to describe subresolution reconnection events (Parker 1983). Soon after, observers used them for small flare-like brightenings (e.g. Lin {\sl et al.,} 1984; Gary {\sl et al.,} 1997). Here we use the term "micro-event" to denote short enhancements of coronal emissions in the energy range of about $10^{24} - 10^{28}$erg. The lower limit is given by current instrumental thresholds observing quiet regions, and the upper limit refers to the smallest events observable typically in an active region. As the observable micro-events in quiet regions are significantly smaller than micro-events in active regions, it is convenient to refer to them {\sl nanoflares} and {\sl microflares}, respectively. Although the difference between nanoflares and microflares first appears just in energy, they occur in different physical environments and their relation needs to be investigated. 

{\sl Active region} small flares in the 3 -- 8 keV range lasting a few minutes have been first observed by HXIS on SMM (Simnett {\it et al.}, 1989). Shimizu (1995) found microflares (active region transient brightenings) in Yohkoh/SXT observations sensitive to 4--7 MK. He reported events having thermal energy contents as low as a few $10^{26}$ergs and a correlation of peak flux with source size (loop length) and density. Watanabe {\it et al.} (1995) have observed the full Sun in the high-temperature Sulfur XV lines using the BCS instrument on board Yohkoh, indicating brightenings in excess of 10 MK. Note in contrast that the thermal energy content of micro-events in the quiet corona is more than two orders of magnitude smaller than in active regions, and temperatures of only 1.0--1.6 MK are reported (Krucker {\it et al.}, 1997, and reviewed by Benz \& Krucker, 2002).

Lin {\it et al.} (1984 and references therein) have discovered small HXR flares produced by $>$20 keV electrons. Comparing the $>$25 keV channel of LAD/CGRO with microflares observed in the 8--13 keV channel of SPEC/CGRO, Lin, Feffer \& Schwartz (2001) reported many low-energy events that have no counterpart at high energies. If interpreted as emission of the same non-thermal electron population, it would suggest that the spectrum steepens or cuts off between 10 and 25 keV. Thus the nature of the low-energy events and their relation to the non-thermal events at high energy remain questionable. It can now be investigated by the RHESSI satellite (Lin {\it et al.}, 2002) at unprecedented spectral and temporal resolution of photons above 3 keV. RHESSI is highly sensitive to the highest temperatures reported for active regions as well as to the lowest energies of non-thermal photons expected from microflares. First RHESSI results on microflares have been presented by Benz \& Grigis (2002) and Krucker {\it et al.} (2002). Here we explore their difference to nanoflares and concentrate on their relevance for heating quiet and active regions.

In the {\sl quiet corona}, small brightenings above the network of the magnetic field were first discovered in deep soft X-ray exposures by Yohkoh/SXT (Krucker {\it et al.} 1997). The events were reported to have a typical thermal energy of $10^{26}$erg, and to occur at a rate of 1200 per hour extrapolated over the whole Sun. Thus they represent a population of ubiquitous dynamic coronal phenomena, quite different from the more sporadic microflares occurring in active regions that are strongly related to solar activity and its cycle. Nanoflares also appear different from X-ray bright points discovered by Skylab that emerge at a rate of 62 per hour averaged over the whole Sun and persist on average for 8 hours (Golub {\it et al.} 1974). The radiative energy loss during the bright point's lifetime is of the order of $10^{28}$ erg. They seem to be the place of frequent nanoflares (Pre\'s \& Phillips 1999). More sensitive measurements of nanoflares became possible with EIT and TRACE. Berghmans, Clette \& Moses (1998) and Benz \& Krucker (1998) found independently that the coronal emission measure observed in EUV iron lines fluctuates locally at time scales of a few minutes in a majority of pixels including even the intracell regions of the quiet corona. At the level of 3 standard deviations, Krucker \& Benz (1998) reported the equivalent of 1.1$\times 10^6$ coronal micro-events per hour over the whole Sun for observations by the Extreme ultraviolet Imaging Telescope (EIT) on SoHO (Delaboudini\`ere {\it et al.} 1995) and noted a nearly linear relation between inferred averaged input power and radiative loss per pixel.

Contrary to micro-events in the corona, localized brightenings in the transition region and chromosphere have been reported for decades. Habbal (1992) and Brkovi\'c {\it et al.} (2000) have pointed out that some of them are associated with a coronal brightening, others are not. Here we concentrate on events that have an impact on the corona. Thus we search for an observable change of the energy density in localized regions of the corona and will consider only coronal emissions in the following.

\section*{ACTIVE REGION MICRO-EVENTS}
\vskip3mm
Figure 1a displays RHESSI full Sun observations at low energies during one orbit. The low peak count rates justify neglecting the pile-up effect of photons in the detectors. The photons have been integrated in one keV bins that we will refer to as channels. The 10--11 and 11--12 keV channels have a lower signal-to-noise ratio because of an instrumental background line feature. Also displayed in Figure 1a are the GOES 8 observations in the low-energy band, extending the energy coverage below the RHESSI range. 

\begin{figure}
\centerline{
\includegraphics[width=175mm]{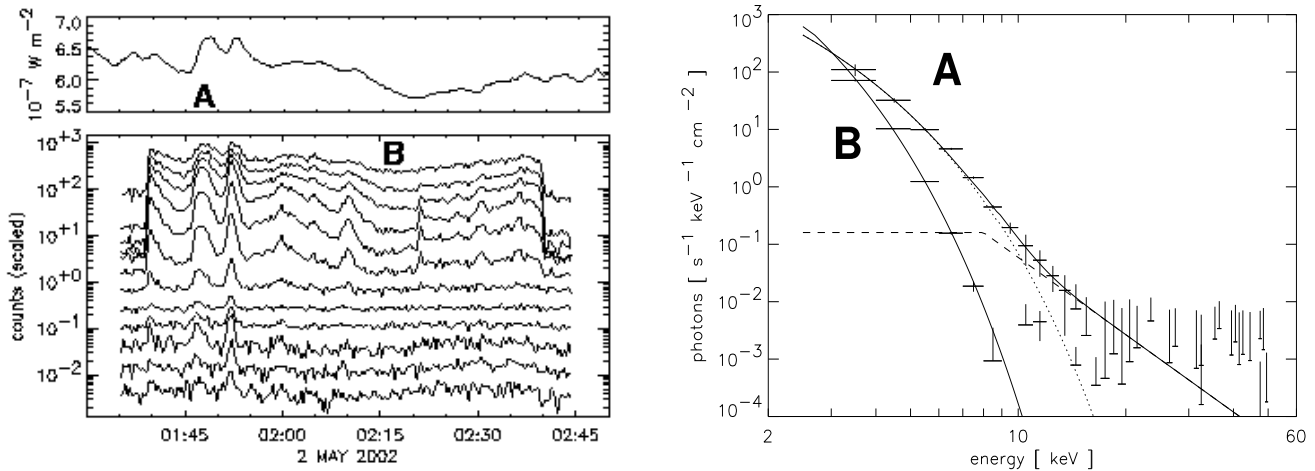}}
\caption[fig1]{{\sl Left top:} Light curve observed by GOES 8 on 2-May-2002 in the 1--8 \AA\ band (1.6--12.4 keV) of soft X-rays. {\sl Left bottom:} RHESSI light curves in one keV channels from 3--4 keV (top) to 14--15 keV (bottom). Each channel is multiplied such that it does not overlap with others. The time resolution is 20 s. Sun light starts at 01:40 UT and ends at 02:40 UT.
{\sl Right:}  Spectrum of event A and non-flare level B (Fig.1a) observed with RHESSI. The event is fitted with a thermal contribution (dotted) and a non-thermal component (dashed) having a break point at 8 keV. The non-flare level is fitted with only a thermal contribution (Benz \& Grigis, 2002).}
\end{figure}

RHESSI observations have been selected at times when the full flux of low-energy photons is available. As in the early part of the mission an attenuator was often placed into the field of view to reduce the number of low-energy photons, we have searched for unattenuated orbits in the months of March, April and May 2002. Uninterrupted intervals at low solar activity and low non-solar background were selected for further analysis. A total of seven intervals satisfied these requirements. The GOES level in all of them is below C, and all flares are below GOES class A9 after subtracting the pre-flare level.

Figure 1a exhibits many events in one of the orbits. Indeed, the flux is never constant. During the total selected observing time of 373 minutes, we found 64 events, thus 10.3 per hour on average in one solar hemisphere. Lin {\it et al.} (1984) reported 10 h$^{-1}$, Lin {\it et al.} (2001) 5.5 h$^{-1}$, and Shimizu (1995) 26 h$^{-1}$, all of them observing during higher levels of solar activity. 

The largest variability in Figure 1a occurs in the channels between 5--9 keV. All events visible in the GOES light curve, being dominated by photons $<$ 3 keV, can also be recognized in the 6--9 keV channels. The event duration decreases continuously from 1.6 keV (GOES) to 15 keV. In most events there is no emission perceivable beyond about 12~keV.

Two spectra are shown in Figure 1b. The data were calibrated in July 2002 using the full detector response matrix. The first spectrum was integrated during event A shown in Figure 1a from 01:46:40 to 01:47:40 UT, the maximum microflare phase at the higher energies. The pre-flare level, taken at the time of minimum flux before the event, was subtracted. At low energies the spectrum fits well with an isothermal plasma at a temperature of 12.1($\pm$ 0.4) MK and an emission measure of 7($\pm 2)\times 10^{45}$cm$^{-3}$. The best fit is shown in Figure 1b, where all normalized residuals are below 0.4. The errors originate mainly from the choices of the pre-flare level and the low-energy break point for non-thermal emission. As the power-law distribution of the non-thermal component must turn over at low energies, it was approximated by a constant below some energy. The quality of the fit is not sensitive to the break energy; best fits are between 6 and 9 keV. In the case of Figure 1b different pre-flare choices changed temperatures in the range of $\pm 0.8$ MK, and different choices for the low-energy break for non-thermal emission produced a range of $\pm 0.7$ MK. 

In the energy range of 10 -- 15 keV, the isothermal model does not fit the observations of microflare A (Fig.1b). A power-law photon distribution was therefore added. Its fitted slope depends on the break point and has an exponent in the range $-4.5 > \gamma > -5.7$. For the choice of the break point at 8 keV, the best fit yields -4.5. The spectrum disappears in the instrumental background beyond 15 keV.

The thermal energy content of an isothermal plasma is

\begin{equation}
E_{th} \approx 3\sqrt{{\cal{M}}V} k_B T
\end{equation}
It is assumed in Eq.(1) that the electron and ion densities are about equal and constant in the volume $V$. $\cal{M}$ is the emission measure, and $k_B$ is the Boltzmann constant. Putting in the observed values, we find for the thermal content of the microflare (index A) and of the hot kernel (index k)

\begin{equation}
E_{th}^A \approx 4.2\times 10^{14}\sqrt{V_A} \ \ \ \ \ {\rm and}\ \ 
E_{th}^{k} \approx 1.2\times 10^{14}\sqrt{V_{k}} \ \ \ \ \ {\rm [erg]}\ \ ,
\end{equation}
Thus the hot kernel would contain less energy than the main component for $V_k < V_A$. 

\begin{figure}
\centerline{
\includegraphics[width=135mm]{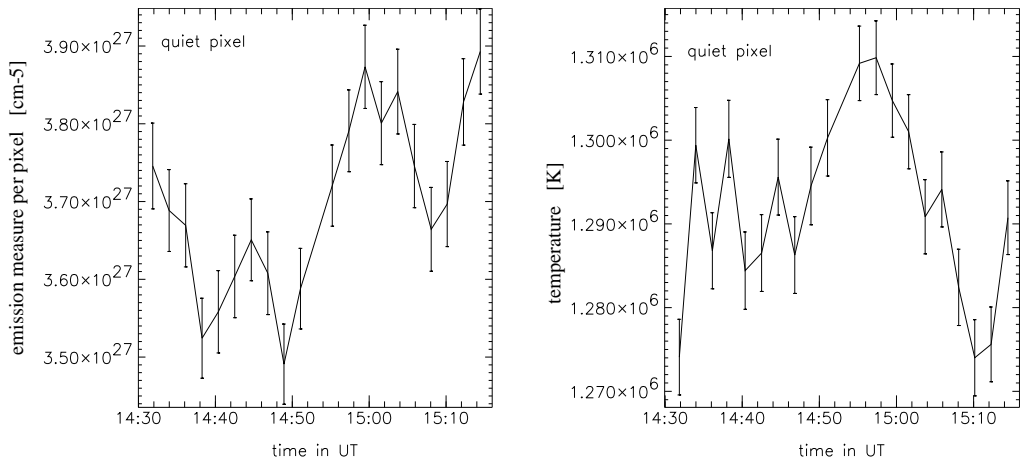}}
\caption[fig2]{{\sl Left:} Coronal mission measure of a random 1900km$\times$1900km pixel in a quiet region of the Sun as determined from EIT high-temperature iron lines. The emission measure is divided by the pixel area $A_p$. The error bars are conservative upper limits as explained in the text. {\sl Right:} The formal temperature evolution of the same pixel is shown. No solar background is subtracted.}
\end{figure}

The spectrum at the time of the lowest count rate in the orbit (marked B in Fig.1a) is also displayed in Figure 1b. It is the time interval, when no microflare was observed and the most recent one had occurred more than 4 minutes before (Fig.1a). For this case, the instrumental background interpolated from the satellite's night time before and after the observations was subtracted. An isothermal model having a temperature of 6.4 MK fits well. No non-thermal or hot kernel component is above the background level. The emission measure of the hemisphere is measured $45\times 10^{45}$cm$^{-3}$, and its thermal energy content is

\begin{equation}
E_{th}^{B} \approx 5.6\times 10^{14}\sqrt{V_B} \ \ \ \ \ {\rm [erg]}\ \ .
\end{equation}
The fact that Eqs.(2) and (3) have similar factors does not mean that $E_{\rm th}^A \approx E_{\rm th}^B$ since probably $V_B \gg V_A$.

Microflare A can be imaged with RHESSI collimators 3 to 9. The microflare took place in NOAA Active Region 9932 and had a nominal FWHM size of 8$''$$\times$11$''$. It is close to the resolution limit, thus $V_A \lapprox 3\times 10^{26}$cm$^3$. Eq.(2) yields an upper limit on the thermal energy content of $7.2\times 10^{27}$erg. A lower limit for the electron density of $4.8\times 10^{9}$cm$^{-3}$ is derived from emission measure and volume assuming a filling factor of unity. 

The emission above about 9 keV peaks generally before the emission at lower energies. The temperature maximum occurs between peak fluxes in 12--15 keV and 3--7 keV. The product $T\sqrt{{\cal M}}$ is proportional to the energy content if the volume remains constant. The product increases initially until about 01:48:30 s, when the 12--15 keV count rate has dropped to the half-maximum level, and then remains approximately constant throughout the microflare.

We conclude from the first RHESSI results that the spectrum of microflares below 9 keV is usually dominated by a thermal component with a temperature of the order of 10 MK. The thermal component can be fitted by temperatures of 6--14 MK and has an average duration of 131($\pm$103) seconds at 7--8 keV. It can be traced up to some 12 keV in the spectrogram. It can be separated from a second component at higher energies, not visible in all events, that peaks simultaneously at all energies, has short duration and occurs during the rise phase of the first component. In 24\% of the cases this enhanced high-energy tail is observed extending up to 15 keV and can be traced down to about 8 keV. The existence of two components excludes the interpretation of one non-thermal electron population having a cut-off distribution in energy.

A hot kernel of thermally emitting plasma also fits the 9--14 keV spectrum, but would produce a drift during cooling that is not observed. Furthermore, this spectral component has shorter duration, occurs in the early phase of the thermal component and vanishes at the peak flux of the thermal component. It resembles the non-thermal component of regular flares in its relation to the first component known as the Neupert effect and confirms an earlier result of Lin {\sl et al.} (1984). The second component thus exhibits the characteristics of thick target bremsstrahlung of an electron beam. Therefore, Benz \& Grigis (2002) have interpreted the high-energy component of microflares as non-thermal emission. 

\begin{figure}
\centerline{
\includegraphics[width=175mm]{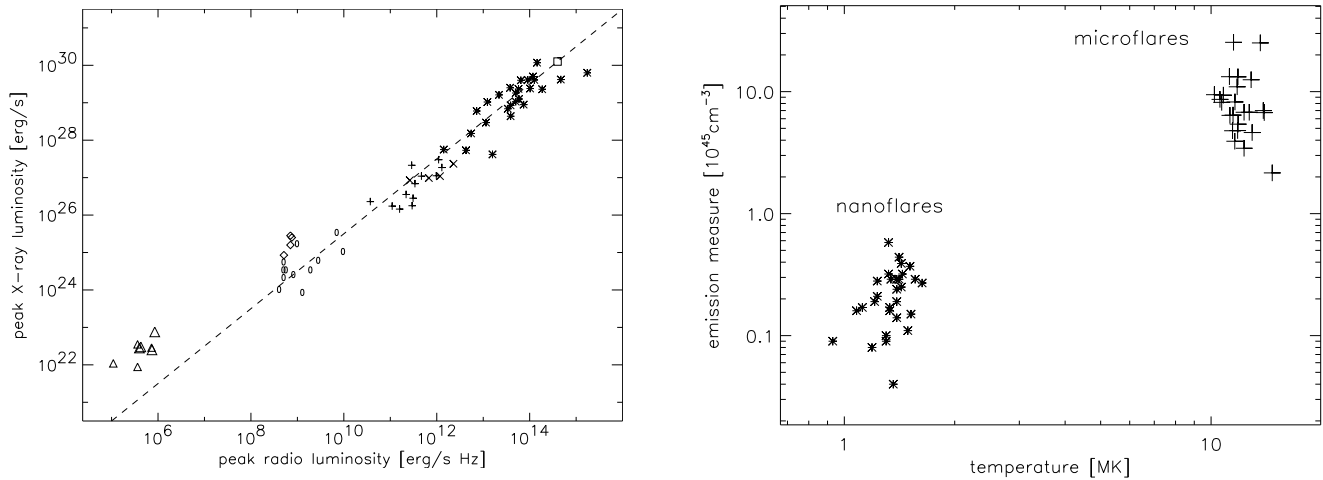}}
\caption[fig3]{{\sl Left:} Radio vs SXR peak flare fluxes from nanoflares (triangles), microflares (diamonds and circles), impulsive flares (plus), gradual flares (times), quiescent stellar emission (asterisks) to a stellar flare (quadrangle)  (from Benz \& Guedel, 1994, and Krucker \& Benz, 2000). {\sl Right:} Emission measure vs. temperature for nanoflares in quiet regions (Benz \& Krucker, 2002, using EIT data) and microflares in active regions (Benz \& Grigis, 2002, using RHESSI data).}
\end{figure}

The similarities of microflares to regular flares suggest identical physical processes and similar plasma conditions. The properties of the thermal and non-thermal components are consistent with the standard flare scenario of precipitating electrons heating cold material to flare temperature. The collisional energy loss of beaming electrons causes bremsstrahlung emission observable in the 9--15 keV range and heats cold material to emit thermally at $<$12 keV. 

The duration increases with decreasing photon energy. The peak count rate defined by cross-correlation is delayed at low energies. The temperature peaks early in the event and then decreases, whereas the emission measure increases throughout the event. The properties are consistent with thermal conduction dominating the evolution. In some of the bigger events, a second component was found in the 11--14 keV range extending down to 8 keV in some cases. The duration is typically 3 times shorter and ends near the peak time of the thermal component consistent with the Neupert effect of regular flares. The two components can be separated and analyzed in detail for the first time. Low-keV measurements allow a reliable estimate of the energy input by microflares necessary to assess their relevance for coronal heating.

\section*{QUIET REGION MICRO-EVENTS}
\vskip3mm
In the following section micro-events in active regions are compared with their counterparts in quiet regions of the corona. The emphasis is on emission lines observed in the two wavelength bands, 171 \AA \ and 195 \AA, including lines of Fe IX/X and Fe XII, respectively, with diagnostic capabilities in the range of the plasma temperatures yielding significant abundances of the two ions. These coronal lines dominate the observed passbands of the EUV spectrum, and their large photon fluxes provide higher sensitivity than previous observations in soft X-rays (Krucker {\it et al.}, 1997). Using lines of different ionization states, the emission measure in a range of temperatures can be derived. The ratio of the emission in the two passbands also defines a line-ratio temperature in the range of about 1.0--1.9 MK (Moses {\it et al.,} 1997). The calculation of these basic quantities employs a spectral line code such as SPEX or CHIANTI.

The line-ratio temperature and emission measure have been determined in Figure 2 for each pixel at each time step from EIT observations. The variability of the quiet corona near the center of the disk is well illustrated by Figure 2. The presented observing run lasted from 14:30 to 15:15 UT on July 12, 1996, when solar activity was almost at the lowest level during the most recent solar minimum. The time resolution is 127.8 s and the pixel size 2.62" (1900 km on the Sun). 

\section*{DISCUSSION}
\vskip3mm

The characteristics of the largest micro-events can be observed in sufficient detail to compare them according to their origin. Nanoflares are accompanied with relatively little radio emission, whereas microflares have a relation between radio (gyro-synchrotron) emission and soft X-ray emission that is similar to regular flares (Fig. 3a).  As the radio emission of microflares was measured at 10 GHz and thus originates by the gyro-synchrotron process, the difference suggests a lower number of relativistic electrons in nanoflares relative to their thermal energy content.  Figure 3b indicates a clear separation between nanoflares and microflares in emission measure ($> 10^6$K) and temperature. Nanoflares are cooler than microflares. Note that the EIT nanoflare observations have a lower limit on temperature of about $1.0\cdot10^6$K, and RHESSI sees microflare only beyond a few $10^6$K. Both observations, however, do not reach the upper limit of the temperature range given by the instruments. Similarly, the instrumental sensitivity puts lower limits to the emission measures. The two sets of data displayed in Figure 3b comprise a similar number of events observed above these limits. Thus the upper right corners of the distributions are not given by the observational methods and can be compared. The extremes in that direction are separated by one order of magnitude in temperature and a factor of 50 in emission measure. Temperatures exceeding the ones of the nanoflares shown in Figure 3b have been reported for coronal bright points (e.g. Pr\'es \& Phillips, 1999), but did not show up in the 42 minute and 7'$\times$7' EIT observation. 

The total energy input can be estimated and compared to the observed requirements, assuming that the smaller events follow the same pattern. A rough estimate of the energy input observed by EIT on the SoHO satellite is of the order of 10\% of the total radiative output in the same region (Benz \& Krucker, 2002). This estimate includes only the part of energy that is deposited in the chromosphere and there creates hot thermal plasma evaporating into the corona. In addition, energy may be released directly into the corona in the form of heating, motion and wave turbulence. Saint-Hilaire \& Benz (2002) have estimated this fraction to be 30\% of the total energy in one well observed regular flare. 

Simulations by Mitra Kraev \& Benz (2000) have shown that the energy input into the corona by unresolved nanoflares produces a slowly variable baseline of emission measure, present all the time. The same effect, however, can also be produced by continuous heating of the corona. A direct coronal energy input by micro-events is equivalent to such continuous heating if the energy is sufficiently distributed across the magnetic field and reaches higher altitudes. A similar consideration can be made for active region heating.

\section*{CONCLUSIONS}

\vskip3mm
The higher resolution of RHESSI has revealed more and smaller flares than previous missions. All microflares occurred in active regions and demonstrate that energy is released at levels far below regular flares. The new results have been contrasted with previous observations of nanoflares in quiet regions and found to be at least quantitatively different. Nanoflares and microflares appear in different ranges of temperature and emission measure. As the instrumental limits prohibit observations at intermediate temperatures, cases with temperature between the two populations cannot be excluded, although temperatures typical for microflares (some 10 MK) have never been reported in the quiet corona. Even if their distribution in emission measure and temperature were the same, the occurrence rates of nanoflares and microflares are so different that they cannot originate from the same population. Future RHESSI analysis can determine the thermal and non-thermal forms of microflare energy with an unprecedented accuracy and will allow to estimate the impact of microflares on the heating of the active region corona.

\vskip4mm
Received:\  29 November 2002,\ \  accepted:\ 7 February 2003.\par
\end{document}